\documentclass[english]{IEEEtran}
\usepackage[T1]{fontenc}
\usepackage{amsmath}
\usepackage{stmaryrd}
\usepackage{graphicx}
\usepackage{multirow}



\usepackage{babel}

\usepackage{cite}
\usepackage{amsmath,amssymb,amsfonts}
\usepackage{algorithmic}
\usepackage{graphicx}
\usepackage{textcomp}
\usepackage{threeparttable}
\newtheorem{theorem}{Theorem}

\def\BibTeX{{\rm B\kern-.05em{\sc i\kern-.025em b}\kern-.08em
    T\kern-.1667em\lower.7ex\hbox{E}\kern-.125emX}}
\begin{document}
% \history{Date of publication xxxx 00, 0000, date of current version xxxx 00, 0000.}
% \doi{10.1109/ACCESS.2017.DOI}

\title{Distributed Coded Modulation Schemes for Multiple Access Relay Channels}

\author{Zihuai Lin, \emph{Senior Member}, 

\thanks{Zihuai Lin is with the School of Electrical and Information Engineering,
The University of Sydney, Australia (e-mail: zihuai.lin@sydney.edu.au).}
}
\maketitle

\begin{abstract}
In this paper, we investigate network nest coded modulation schemes for multiple access relay channels.
The performance of the distributed systems which are based on
distributed convolutional codes with network coded
modulation is presented. An analytical upper bound on bit error
probability performance for the studied distributed systems with
Maximum Likelihood Sequence Detection (MLSD) is derived. The
constructed bounds for the investigated systems are shown to be
asymptotically tight for increasing channel Signal-to-Noise Ratio
(SNR) values. 
\end{abstract}

% \begin{keywords}
% distributed coding, network coded modulation, multiple access relay channel, nested codes
% \end{keywords}

% \titlepgskip=-15pt

% \maketitle

\section{Introduction}
\label{sec:introduction}
%\PARstart{D}{
Distributed coding, as a special channel coding strategy developed
for cooperative communication
networks~\cite{Erkip},~\cite{Hunter}, attracted large attentions
recently. The distributed codes construction concept has been
applied on conventional channel coding to form such as distributed
turbo codes~\cite{DTC}, %distributed space-time codes~\cite{DSPT}
 distributed low-density parity-check (LDPC)
codes~\cite{DLDPC} and distributed rateless codes \cite{DC1,DC2,DC3,DC4,DC5,DC6,DC7,DC8,DC9,DC10,DC11,DC12,DC13}. These published results show that the proposed
schemes can improve the transmission reliability over
point-to-point wireless communication channels.

The distributed coding schemes discussed above are developed for
small-scale unicast relay networks, in which the messages are sent
from a single source to a single destination through
single/multi-hop relays. In this work, we consider a scenario that
multiple source nodes communicate with multiple destination nodes via a Relay Node
(RN). A classical way to pass such kind of information is through
routing, where the relay nodes simply store and forward the
received packets to the destination.

In ~\cite{networkcoding}, a network coding (NC) approach is proposed
to replace routing. In NC, the relay nodes are allowed to encode
the packets received from multiple source nodes. The combined
information is subsequently sent to the destination. It has been
shown in~\cite{NC1,NC2,NC3,NC4,NC5,NC6,NC7,NC8,NC9} that compared with traditional routing, NC
can enhance the network capacity and throughput.

For the existing research in distributed network-channel codes (DNCC)
design, many open questions in the design and
implementation of distributed codes still have not been addressed.
In this work, we develop the design of 
distributed codes for uplink transmissions in cellular systems and analyze the code performance based on the
framework of uplink cellular systems.

The main contributions of this work are the propose of a distributed physical layer network coded system,
the derivation of the analytical upper bound on the
error probability. %and the propose of an iterative decoding approach 
%for the interleaved distributed coded system.

The outline of the paper is given below. The system model is described in Section II. The decoding algorithm is depicted in Section III. Performance analysis in terms of the upper bound on the error probability is given in Section IV. Code search results are given in Section V. In Section VI, numerical and simulation results are presented. Finally, conclusions are drawn in Section VII.

\section{System Model}

We consider a wireless network  with one relay node and a set of source and destination nodes. The source nodes send data packets to the destination nodes via the relay node. Let us denote by $\mathfrak{S}$ the set of source nodes and by $\mathfrak{T}$ the set of destination nodes. A connection between a source node $S$ and a destination node $T$ is established through the relay node $R$, where $S \in \mathfrak{S}$ and $T \in \mathfrak{T}$.
The system can be modeled with a directed hypergraph $\mathfrak{G}=(\mathfrak{N},\mathfrak{E})$ where $\mathfrak{N}$ is the set of nodes, ${\mathfrak{N}=\mathfrak{S} \bigcup \mathfrak{T}} \bigcup R$ and $\mathfrak{E}$ is a set of hyperedges. %which is defined as a connection between any number of nodes of a hypergraph.
A hyperedge $(n,\mathfrak{D})$ consists of the directed edges between a start node $n$ and a set of end nodes $\mathfrak{D}$, where $n \in \mathfrak{N}$, $\mathfrak{D}\subset \mathfrak{N}$ and $\mathfrak{D}$ is non-empty. A hyperedge $(n,\mathfrak{D})$ represents broadcasting links between node $n$ and the nodes in $\mathfrak{D}$.

An example of this scenario is illustrated in Fig. 1, in which four source nodes transmit
the information packets to four destination nodes via the relay node. In this case,
${\mathfrak{S}}=\{A,C,D,F\}$ and ${\mathfrak{T}}=\{B,C,E,F\}$.  Denote by ${\mathbf{I}}_A$, ${\mathbf{I}}_C$, ${\mathbf{I}}_D$, ${\mathbf{I}}_F$ the packets
originated from the source node $A$, $C$, $D$, $F$, respectively. We assume that
all the source nodes have packets to transmit and all the packets have equal length
of $\kappa$ bits. In the concept of network coding, the packets received by the relay node
will be linearly combined over a finite field GF($2$) and then broadcast to all the
destination nodes. With the assumption that all the destination nodes have the
side information from the neighboring source nodes, the destination nodes
can successfully decode the packets which are intended to them. For example, if
the destination node $B$ knows the packets transmitted from source nodes $C$,
$A$, $F$, then it can successfully decode the packet ${\mathbf{I}}_D$. In practical systems,
however, the complete side information from the source nodes may not be
available. For example, node $B$ may only have side information from the
neighbor nodes $A$ and $C$, in this case the node $B$ cannot successfully
decode the packet ${\mathbf{I}}_D$.
\setlength{\unitlength}{0.5mm}
\begin{figure}%[!t]%[H]
%\normalsize
\begin{picture}(00,80)(20.5,50)
\linethickness{0.5pt}

\put(78,110){\circle{7}} \put(78,110){\makebox(0,0){$A$}}

\put(118,110){\circle{7}} \put(118,110){\makebox(0,0){$B$}}

\put(98,78){\circle{8}}\put(98,78){\circle{7}}
\put(98,78){\makebox(0,0){$R$}}
\put(87,110){\makebox(0,0){${\mathbf{I}}_A$}}
\put(59,87){\makebox(0,0){${\mathbf{I}}_F$}}
\put(70,50){\makebox(0,0){${\mathbf{I}}_D$}}
\put(138,71){\makebox(0,0){${\mathbf{I}}_C$}}

\put(138,78){\circle{7}} \put(138,78){\makebox(0,0){$C$}}

\put(58,78){\circle{7}} \put(58,78){\makebox(0,0){$F$}}

\put(78,46){\circle{7}} \put(78,46){\makebox(0,0){$D$}}

\put(118,46){\circle{7}} \put(118,46){\makebox(0,0){$E$}}

\put(61.5,79){\vector(1,0){33}} %\put(45.5,112){\vector(2,-3){32}}
\put(94.5,77){\vector(-1,0){33}}
\put(101.5,79){\vector(1,0){33}} %\put(45.5,112){\vector(2,-3){32}}
\put(134.5,77){\vector(-1,0){33}}
\put(81,49){\vector(2,3){16}} %\put(45.5,112){\vector(2,-3){32}}
\put(99.5,80){\vector(2,3){18}} %\put(121.5,78){\vector(1,1){33}}

\put(100.5,76){\vector(2,-3){17}}
\put(79.5,106.5){\vector(2,-3){17}}

\end{picture}
\vspace{5ex}
\caption{A network coding group with four source nodes and four destination nodes and one relay node.} \label{fig:OneStepTrans}
%\hrulefill
\end{figure}
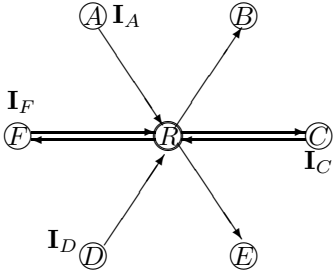

In this work, we consider the system that
the information blocks, e.g., ${\mathbf{I}}_A$, are encoded at the source
node $A$ before its transmission to the relay node. The received
codewords from all the source nodes are linearly combined using the XOR operation at the relay node and then
broadcast to all the destination nodes. Assume error free transmission from the source nodes to the relay node. Then the codeword generated at the relay node is
\begin{equation}
{{\mathbf{C}}}={\mathbf{I}}_A{\mathbf{G}}_A\oplus {\mathbf{I}}_C{\mathbf{G}}_C \oplus {\mathbf{I}}_D{\mathbf{G}}_D \oplus {\mathbf{I}}_F{\mathbf{G}}_F,
\end{equation}
where ${\mathbf{G}}_A$, ${\mathbf{G}}_C$,${\mathbf{G}}_D$,${\mathbf{G}}_F$ are the generator matrices for the corresponding source nodes with a code rate of $\kappa/\eta N_T$, where $N_T$ is the cardinality of the set of destination nodes $\mathfrak{T}$.
Each destination node
decodes the received codeword with its own side information. Let ${\bf{\Theta}}_\omega$ be the side information for the destination node $\omega$, $\omega\in {\mathfrak{T}}$ and $\hat{{\mathbf{C}}}$ be the received codeword after the demodulation. At node $\omega$, the received codeword is combined with the side information,
\begin{equation}
\tilde{\mathbf{C}}_\omega=\hat{\mathbf{C}}\oplus {\bf{\Theta}}_\omega.
\end{equation}
$\tilde{\mathbf{C}}_\omega$ denotes the received mixed codewords which are unknown to the node $\omega$. It represents a corrupted version of the codeword with a code rate $(N_T-|{\bf{\Theta}}_\omega|)\kappa/(\eta N_T)$, where $|{\bf{\Theta}}_\omega|$ is the length of the set of the destination nodes which are known by node $\omega$. The receiver at node $\omega$ can then decode $\tilde{\mathbf{C}}_\omega$ with the corresponding code rate.
Therefore, a single codeword ${{\mathbf{C}}}$ can be  interpreted differently by different destination nodes. This is also the basic idea of nested codes. For more detailed description of nested codes, please refer to \cite{Frenger,Ma-nest}.

For a special case of the above mentioned system, we consider a wireless cellular uplink transmission system, in
which a number of MTs send data packets to a
BS via a RN. %The source MT can collaborate with
%other MTs for cooperative transmission.
We assume that there are
direct links between the source MTs and the RN and between the RN and the BS, but no
direct links between the BS and the MTs. %The discussion of the
%assumption of the direct links between the MTs is given in
%\cite{winner-p2p}.
The transmission can be separated into two phases. In the first phase, the source MTs first encode the
information packets then broadcast to the RN. In the second phase, the RN decodes the received codewords from the source MTs, then the codewords will be
re-encoded, modulated and broadcasted to the BS. 

The system model is illustrated in Fig. 2, in which a number of $S$ source MTs
%cooperating with a cooperative MT
transmit information packets to
a BS via a RN. The $i$th source MT uses the
encoder $C_i$ with a code rate of $1/n_i$. We assume that
all the source nodes have packets to transmit and all the packets have equal length
of $\kappa$ bits. In the concept of network coding, the packets received by the relay node
will be linearly combined over a finite field GF($2$) and then broadcast to all the
destination nodes.

\setlength{\unitlength}{0.5mm}
\begin{figure}%[!t]%[H]
%\normalsize
%\centering
\begin{picture}(00,80)(20.5,50)
\linethickness{0.5pt}

\put(78,110){\circle{7}} \put(78,110){\makebox(0,0){$A$}}

%\put(118,110){\circle{7}} \put(118,110){\makebox(0,0){$B$}}

\put(98,78){\circle{8}}\put(98,78){\circle{7}}
\put(98,78){\makebox(0,0){$R$}} \put(100,86){\makebox(0,0){RN}}

\put(87,120){\makebox(0,0){Source
MT 1}}\put(87,110){\makebox(0,0){$\mathbf{C}_1$}}
%\put(59,87){\makebox(0,0){${\mathbf{I}}_F$}}
\put(78,90){\makebox(0,0){$\vdots$}}
\put(78,70){\makebox(0,0){$\vdots$}}
\put(70,38){\makebox(0,0){Source MT S}}
\put(70,50){\makebox(0,0){$\mathbf{C}_S$}}
\put(138,71){\makebox(0,0){${\mathbf{BS}}$}}

\put(138,78){\circle{7}} \put(138,78){\makebox(0,0){$C$}}

%\put(58,78){\circle{7}} \put(58,78){\makebox(0,0){$F$}}

\put(78,46){\circle{7}} \put(78,46){\makebox(0,0){$B$}}

%\put(118,46){\circle{7}} \put(118,46){\makebox(0,0){$E$}}

%\put(61.5,79){\vector(1,0){33}} %\put(45.5,112){\vector(2,-3){32}}
%\put(94.5,77){\vector(-1,0){33}}
\put(101.5,79){\vector(1,0){33}} %\put(45.5,112){\vector(2,-3){32}}
%\put(134.5,77){\vector(-1,0){33}}
\put(81,49){\vector(2,3){16}} %\put(45.5,112){\vector(2,-3){32}}
%\put(99.5,80){\vector(2,3){18}} %\put(121.5,78){\vector(1,1){33}}

%\put(100.5,76){\vector(2,-3){17}}
\put(79.5,106.5){\vector(2,-3){17}}

\end{picture}
\vspace{5ex} \caption{An uplink transmission system with two
source MTs, a Relay node and a BS.}
\label{fig:OneStepTrans}
%\hrulefill
\end{figure}
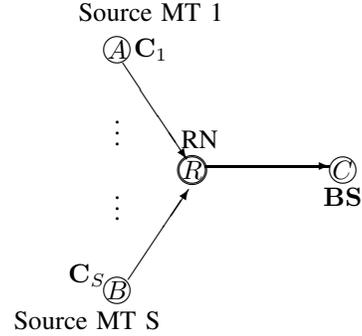

In this work, the transmitted information of each source node is encoded with
different linearly independent generators and then forwarded
to the relay node. Since the different transmitted information will
be nested at the relay node with XOR operation, the rate of nested packets
will be easily higher than one if we encode source information with a high
rate channel code. For example, if we choose a code with rate $R=k/n$
to encode different source information separately and XOR them, the nested rate will be
$R_{nested}=S\cdot k/n$ ($S$ denotes the number of source nodes), which easily achieves a much higher rate than one.
However, such a high rate can only be decoded by perfect prior knowledge at
the destination nodes. % \cite{06Xiao}.

Alternatively, select a low rate $R/S=k/(nS)$ code to encode information
at different source node, and then perform XOR operation at
the relay node, mathematically,
\begin{eqnarray}\label{equ nested os}
%\begin{split}
c_{nested}&={\mathbf{i}}_{1}{\mathbf{G}}_{1}\oplus {\mathbf{i}}_{2}{\mathbf{G}}2_{2}\oplus\cdots\oplus {\mathbf{i}}_{S}{\mathbf{G}}_{S}\\
&=[{\mathbf{i}}_{1},{\mathbf{i}}_{2},\cdots,{\mathbf{i}}_{S}][{\mathbf{G}}_{1},{\mathbf{G}}_{2},\cdots,{\mathbf{G}}_{S}]^{T},
%\end{split}
\end{eqnarray}
where ${\mathbf{G}}_{1},{\mathbf{G}}_{2,}\cdots,{\mathbf{G}}_{S}$ are mutually linearly independent generators corresponding to
rate $k/(nS)$ codes. $\oplus$ denotes XOR operation.
%Because different information encode with different linearly independent generator matrices, we can get multiple interpretations at different destination nodes.
At the relay, the output codewords $c_{nested}$ are then fed into a Network Encoder (NE) and modulated with a
memoryless modulator (MM). By properly choosing the NE and the MM,
we can form a class of digital coded modulation scheme. To simplify the study, here we borrow the idea of coded digital phase modulation scheme, such as CPFSK \cite{proakis}, in which the NE is a Recursive Systematic Convolutional (RSC) encoder. It was shown in \cite{Rimoldi} that a CPFSK scheme can be decomposed into a concatenation of a RSC and a MM.
CPFSK has the advantage of creating a differentially encoded waveform stream and is further attractive due to its good spectral properties, which has been widely used in many communication systems, e.g. digital video broadcasting (DVB) \cite{DVB_1}, satellite communication networks \cite{Sat_1, Sat_2, Sat_3}, etc. The generated
waveform from the digital phase modulator is transmitted over the channel.
The channel is assumed to be the AWGN channel. 

\section{Decoding with Nested Coded Modulation at the Destination Node}

The received passband signal at the destination node can be expressed as,
$r(t,\boldsymbol{\mu})=s(t,\boldsymbol{\mu})+n(t)$, where $n(t)$ is
a zero mean
Gaussian random process. % with double sided power spectral density
$s(t,\boldsymbol{\mu})$ is the
transmitted signal from the relay with binary
information symbols $\mu \in \{0,1\}$, which can be written
as \cite{Rimoldi}
%\vspace{-1ex}
\begin{equation}\label{cpmtrans}%\vspace{-1ex}
  %s(t,\boldsymbol{\alpha})=\sqrt{\frac{2E}{T}}\cos(\omega_0t+\phi(t,\boldsymbol{\alpha}))
  s(t,\boldsymbol{\mu})=\textup{Re}\left\{s_b(t,\boldsymbol{\mu})e^{j(2{\pi}f_1t+\phi_0)}\right\}
%s(t,\textbf{U})=\textup{Re}\left\{\sqrt{\frac{2E}{T}}e^{j(2{\pi}f_1t+\overline{\psi}(t,\textbf{U})+\phi_0)}\right\}
\end{equation}
where
%\begin{equation}\label{cpmtrans}
%  %s(t,\boldsymbol{\alpha})=\sqrt{\frac{2E}{T}}\cos(\omega_0t+\phi(t,\boldsymbol{\alpha}))
$s_b(t,\boldsymbol{\mu})=\sqrt{\frac{2E_s}{T}}e^{j\overline{\psi}(t,\boldsymbol{\mu})}$
%\end{equation}
is the complex baseband equivalent signal \cite{w&j}.
$f_1=f_c-h/2T$ is a shift of the carrier frequency $f_c$,
$\phi_0$ is the initial phase of the carrier and
$\boldsymbol{\mu}$ is the data symbol sequence.
%$\textbf{U}={U}_0,{U}_1,\cdots$.
$E_s$ and $T$ are the symbol energy
%$E=\log_2M{\cdot}E_b$, where $E_b$ is the bit energ
and the symbol interval duration, respectively.

The \textit{tilted information carrying phase} \footnote{In
\cite{Rimoldi} %, Rimoldi used the concept of tilted phase
representation of CPM.}
during symbol interval $n$ $(t=\tau+nT)$
is given by (\ref{phase}),
\begin{figure*}[!t]
\normalsize %%%%%%\setcounter{MYtempeqncnt}{\value{equation}}
\begin{eqnarray}\label{phase}%\vspace{-1ex}
\overline{\psi}(\tau+nT,\boldsymbol{\mu})&=&
\emph{R}_{2\pi}\left\{
2{\pi}h\emph{R}_P\left\{\sum_{i=0}^{n-L}\mu_i \right\}+
4{\pi}h\sum_{i=0}^{L-1}\mu_{n-i}q(\tau+iT) +W(\tau)
%\nonumber\\
\right\},
%\nonumber\\
%\hspace{1mm}
%&&
\hspace{2mm} 0 \leq \tau < T \nonumber\\
\end{eqnarray}
 \hrulefill
\end{figure*}
where $\emph{R}_x\{\cdot\}$ is the modulo $x$ operator and $W(\tau)$ is given by (\ref{eq:wtau}), which
\begin{figure*}[!t]
%\normalsize \setcounter{MYtempeqncnt}{\value{equation}}
\setcounter{equation}{4}
\begin{eqnarray}\label{eq:wtau}%\vspace{-2ex}
W(\tau)={\pi}h(M-1)\frac{\tau}{T}-2{\pi}h(M-1)\sum_{i=0}^{L-1}q(\tau+iT) %\nonumber\\
%&&
+{\pi}h(M-1)(L-1), \hspace{2mm}
 0 \leq \tau < T \nonumber\\
\end{eqnarray}
 \hrulefill
\end{figure*}
%$W(\tau)$
represents the data-independent terms. The phase response
$q(\cdot)$ is found from the frequency pulse $g(t)$ according to
%\begin{eqnarray}\label{phaseresponse}\vspace{-2ex}
 $q(t)=\int_{-\infty}^{t}g(\tau)d\tau$. %; & &-\infty < t < \infty
%\end{eqnarray}
%where $g(t)$ is the frequency response of duration $L$ symbol
%intervals.
$L$ is the length of the frequency response. For CPFSK, $L=1$ and the frequency pulse
used in this work are Rectangular (REC) pulse %and Raised Cosine
%(RC) pulse
defined as:
\\
LREC $\hspace{3ex}$ $g(t)=\frac{1}{2T}, \hspace{3ex} 0\leq t \leq
T$\\
%
%LRC  $\hspace{2ex}$  $g(t)=\frac{1}{2LT}[1-\cos(\frac{2\pi
%t}{LT})], \hspace{2ex} 0\leq t \leq LT$.

%$N_0/2$.
The equivalent complex baseband continuous-time signal
can be written as \addtocounter{equation}{0}
\begin{equation}\label{cpm}%\vspace{-2ex}
  r_b(t,\boldsymbol{\mu})=s_b(t,\boldsymbol{\mu})+n_b(t),
\end{equation}
where $n_b(t)$ is a complex baseband representation of the
additive white Gaussian random process having zero mean %and the
%autocorrelation function
%$E[n_b(t)n_b(t+\tau)]={2N_0}\delta(\tau)$.
%and
double-sided power spectral density $2N_0$ \cite{w&j}.

In each symbol interval, a set of possible CPM sequences consists
of $P{\cdot}M^L$ various complex valued signals. Thus, a bank of
$P{\cdot}M^L$ complex valued filters, matched to those signals and
sampled once every symbol interval produces a sufficient
statistics, since they form a basis for signal space \cite{w&j}. A
component of the vector $\textbf{r}_n$ representing the sampled
outputs at symbol interval $n$, can be calculated by %\vspace{-1ex}
\begin{equation}\label{eq:MF}%\vspace{-1ex}
r_{i,n}=\int_{(n-1)T}^{nT}r(t,\boldsymbol{\mu})e^{j\overline{\psi}(t,\tilde{\boldsymbol{\mu}}_i)}dt,
%\hspace{2ex} i\in\{1,2,\ldots,P{\cdot}M^L\},
\end{equation}
%for $\forall i\in\{1,2,\ldots,P{\cdot}M^L\}$.
where $i\in\{1,2,\ldots,P{\cdot}M^L\}$ and $\tilde{\boldsymbol{\mu}}_i$ is the $i$th hypothesis %tentative
sequence offered by the receiver. All the $P{\cdot}M^L$ various
complex signals must be generated by using various
$\tilde{\boldsymbol{\mu}}$ s and the index $i$ on the right hand
side of (\ref{eq:MF}) is intended to reflect this.

Sufficient statistics can be produced in many other ways,
corresponding to choosing other basis functions for the signal
space. Changing from one to another is done by a linear
transformation. Given the above choice, signal space basis may not be
orthogonal. There can be linear deterministic dependencies among
the components of $\textbf{r}_n$. Hence $\textbf{r}_n$ has a
non-diagonal covariance matrix $\bf{\Lambda}$.

We denote by $\boldsymbol{\chi}_n$ the expectation of
$\textbf{r}_n$. Because of the Gaussian channel, $\textbf{r}_n$ is
a Gaussian random vector. The joint Probability Density Function
(PDF),
conditioned on the CPE
output vector $\boldsymbol{\nu}_n$, is \cite{LinZihuai_PhDthesis} %\vspace{-1ex}
\begin{equation}\label{eq:CH2_cpmmetric}%\vspace{-1ex}
p(\textbf{r}_n|\boldsymbol{\nu}_n)\propto
\exp{\{-(\textbf{r}_n-\boldsymbol{\chi}_n)^H\bf{\Lambda}^{-1}(\textbf{r}_n-\boldsymbol{\chi}_n)\}}
\end{equation}
where $(\cdot)^H$ denotes Hermitian transpose. A more detailed
description of CPM can be found in \cite{LinZihuai_PhDthesis,ringCCCPM1,ringCCCPM2,IT-CPM,TCOM-CPM1,TCOM-CPM2,ETT-CPM,IET-CPM}.

Based on (\ref{eq:CH2_cpmmetric}), we can use maximum a posterior (MAP) algorithm to decode the received CPFSK signal as described in \cite{LinZihuai_PhDthesis,improvedlowbd,ISIT05RTCQCPM,tcqcpmICC}. Please note that since the generators at the source nodes and the CPFSK modulator formed a super-trellis at the destination nodes, we can employ the MAP algorithm over the super-trellis to decode the transmitted signal for each source node.

At different destination nodes, after the demodulation and decoding process,
we get output signal $\hat{\tau}=c_{nested-os}+e$, where $e$
is the binary error pattern. At the destination node $d_{i}$, $\hat{\tau}$ can
be interpreted as \cite{06Xiao}

\begin{equation}
\hat{\tau}=\bigoplus_{l\notin\kappa_{d_{i}}}i_{l}G_{l}\oplus\bigoplus_{l'\in\kappa_{d_{i}}}i_{l'}G_{l'}+e,
\end{equation}
where $\kappa_{d_{i}}$ denotes the indices of the information
prior known to the destination node $d_{i}$. Part $\bigoplus_{l'\in\kappa_{d_{i}}}i_{l'}G_{l'}$
represents the known information to $d_{i}$, which
can be canceled by XOR operation. Then,

\begin{equation}
\hat{\tau}_{d_{i}}=\hat{\tau}\oplus\bigoplus_{l'\in\kappa_{d_{i}}}i_{l'}G_{l'}=\bigoplus_{l\notin\kappa_{d_{i}}}i_{l}G_{l}+e.
\end{equation}

The remained part of $\bigoplus_{l\notin\kappa_{d_{i}}}i_{l}G_{l}+e$ indicates the
combined unknown nested packets to the destination node $d_{i}$.
To separate the desired information, we need to employ the feature of nested codes. Because of the nested approach in Eq. (\ref{equ nested os}), where $[G_{1},G_{2},\cdots G_{j},\cdots,G_{N_{d}}]^{T}$ can be regarded
as a new ``stacking'' generator matrices, it is possible to
interpret $i_{l}\left(l\notin\kappa_{d_{i}}\right)$ from $\bigoplus_{l\notin\kappa_{d_{i}}}i_{l}G_{l}+e$
with corresponding ``stacking'' generator matrices.

The LLR of the $i$-th bit in $c_{u}$ can be computed as Eq. (\ref{LLr}).
\begin{figure*}[!t]
\begin{equation}\label{LLr}
%\begin{split}
 %&
 L_{c_{u}(i)} \triangleq \frac{\mathbb{P}\mathrm{r}[c_{u}(i) = 0]}{\mathbb{P}\mathrm{r}[c_{u}(i) = 1]}\\
%&
=\left\{
   \begin{aligned}
   L_{c_{nested}^{OS}(i)} &= \log \frac{\mathbb{P}\mathrm{r}[c_{u} \oplus c_{c}(i) = 0]}{\mathbb{P}\mathrm{r}[c_{u}\oplus c_{c}(i) =1 ]}  ~~~\textrm{if}~ c_{c} = 0 \\
   - L_{c_{nested}^{OS}(i)} &= \log \frac{\mathbb{P}\mathrm{r}[c_{u} \oplus c_{c}(i) = 1]}{\mathbb{P}\mathrm{r}[c_{u}\oplus c_{c}(i) =0 ]}  ~~~\textrm{if}~ c_{c} = 1\\
   \end{aligned}
   \right..
%\end{split}
\end{equation}
\end{figure*}
There is no information lost through the above cancelation operation, because it only changes the sign of the LLR. Then we get the calculated LLR $L_{c_{u}}$, which is the estimated soft information of unknown packets to receiver $d$. Based on the linearly independent feature of nested codes, we can separate all the desired information at different destination nodes. Particularly, regarding $[G_{1},G_{2},\cdots G_{j},\cdots,G_{|\mathfrak{S}_{e}|}]^{T}$ from Eq.  (\ref{equ nested os}) as a  new ``stacking'' generator matrix, we can multiple interpret $i_{l}\left(l\notin\kappa_{d}\right)$ from $L_{c_{u}}$.

%For a rate $1/n$ convolutional encoder, the
%generator polynomial matrix is \cite{viterbi} %\cite{S.Lin} %usually in the form of
%%\begin{equation}
%$G_N(D)=
%\begin{bmatrix}
%G^0{(D)}\\
%\vdots\\
%G^{n-1}{(D)}\\
%\end{bmatrix}$,
%%\end{equation}
%where $G^i(D)=g_0^i+g_1^iD+\cdots+g_m^iD^m$ for $i
%\in \{0,\cdots,n-1\}$. The coefficients $g_j^i$ for $0 \leq j \leq m$
%belong to the set $\{0,1\}$. A high rate Punctured
%Convolutional Code (PCC) can be obtained by puncturing a parent
%$1/n$ binary convolutional code. The operation of puncturing some
%coded symbols is implemented by using an $(n\times p)$ puncturing
%matrix, $P_{mat}$, where $p$ is the puncturing period. Let $s$ be
%the total number of transmitted bits during a puncturing period
%$p$, the coding rate of a PCC is $r=p/s$.  For two punctured convolutional codes obtained from the same parent code, we call them rate compatible if the higher rate code is obtain by puncturing the lower rate code according to the rate compatibility criterion. This criterion requires that all the lower rate code make use of all code bits of the high rate code, plus one or more additional bits. In other words, the puncturing matrix of the high rate code is embedded into the puncturing matrix of the lower rate code.

\section{Analytical bounds on the bit error probability for the multiple access relay channels}

In this work, we will develop analytical upper bounds on bit error probability for the investigated network coded system under MLSD.
The different convolutional encoders at different source nodes and the NE at the RN
constitute a super-trellis encoder. At the destination nodes, we can develop a ML decoder for the super-trellis encoder. %we adopt the first decoding methodology,
%\textit{i.e.},
%while the decoding is based on a varying trellis structure.
Let $P_b^u$ be the bit error probability for the links from all source nodes to the relay node. Denote by $P_b^d$ the bit error probability for the link from the relay node to the destination node which has no side information from any source nodes. Then the bit error probability from all source nodes to this destination node is
\begin{equation}\label{eq:BERoverall}
P_b=1-(1-P_b^u)(1-P_b^d).
\end{equation}
Let $N_s$ be the cardinality of the set of source nodes $\mathfrak{S}$. We assume that at each source node, the source information is firstly convolutional encoded, then modulated with BPSK and transmitted over an AWGN channel. We further assume that the sequences from different source node are transmitted over orthogonal channels, so that they do not interfere with each other.
Under the above assumptions, the $P_b^u$ can be expressed as
\begin{equation}\label{eq:BERoverall_2}
P_b^u=1-\prod_{i=1}^{N_s}(1-P_b^{ue}(i)),
\end{equation}
where $P_b^{ue}(i)$ is the bit error probability for the link from the $i$th source node to the relay node. Suppose the code rate at the $i$th source node is $r_i$, and the bit energy is $E_b^i$, then $P_b^{ue}(i)=Q(\sqrt{r_iE_b^i/N_0})$. Thus, $P_b^u$ can be expressed in a closed form.

Now let us look at $P_b^d$. Exact expression of $P_b^d$ is hard to find, however, we can give an upper bound on $P_b^d$ under MLSD. The state ${\boldsymbol{\sigma}}_j$ of the super-trellis encoder
at discrete time $j$ is defined as
$({\boldsymbol{\sigma}}^{cc}_j,{\boldsymbol{\sigma}}^{ne}_j)$,
where ${\boldsymbol{\sigma}}^{cc}_j$ and
${\boldsymbol{\sigma}}^{ne}_j$ denote the state of the Joint
Distributed Convolutional Encoder (JDCE) and the state of RSC at
discrete time $j$, respectively. For a JDCE having $m$ memory
elements, and a CPFSK scheme, the total number of states is
$2^{m+1}$. %For a trellis coded CPM system, the
The state transition
${\boldsymbol{\sigma}}_j\rightarrow{\boldsymbol{\sigma}}_{j+1}$ is
determined by the input of the source nodes. Associated with
this transition is also the input symbol $\mu\in\{0,1\}$ of the
NE, i.e. the RSC encoder, and the mean vector which is obtained by letting the
transmitted waveform pass through a bank of complex filters
which are matched to the transmitted signals
\cite{LinZihuai_PhDthesis}.

%The APP decoding algorithm \cite{BCJR} for an un-interleaved coded CPM system %developed for ring convolutional coded
%%CPM systems in \cite{ISIT05RTCQCPM} can be used for concatenated
%%PRCC/CPM systems with some modification. The APP algorithm
%computes the APP $Pr(U_k=\mu|\textbf{r}_1^\ell)$ of an
%input symbol $\mu$ of the CPE at symbol interval $k$ conditioned
%on a sufficient statistic
%$\textbf{r}_1^\ell=(\textbf{r}_1,\cdots,\textbf{r}_\ell)$ based on
%channel observations $r(t,\boldsymbol{\mu})$, where $\ell$ is the
%length of the input data sequence to the CPE.
%
%
%\subsection{Asymptotical Performance of PRCC/CPM}
%In \cite{tcqcpmICC}, we developed an upper bound on the channel
%distortion %\footnote{The channel distortion is the signal
%%distortion caused by the noisy channel, further explained in
%%\cite{tcqcpmICC}.}
%for a combined Trellis Coded Quantizer (TCQ)
%with binary convolutional coded CPM system under MLSD.
%In this work, we will derive an upper bound on the symbol error
%probability for the punctured trellis coded CPM system under MLSD.
%The bound
%is based on the transfer function technique \cite{SER-bound}. %,
%which is modified and generalized to time variant trellis. This, in turn, is based upon the union bound. %It is
%shown in \cite{tcqcpmICC} that the developed analytical bounds are
%consistent with the simulation results.
%This method can also be
%applied to the concatenated PRCC/CPM.
In this work, we assume all the source nodes
have the same transmit power. Let $E_b$ be the information bit
energy and $N_0/2$ be the double sided power spectral density of
the additive white Gaussian noise. % Furthermore, we employ the hybrid type II ARQ scheme \cite{Linzihuai_masterthesis} in the links between MTs and the RN to ensure that the codewords output from the joint decoder at the RN can be estimated without any error. This can be done by appending a Cyclic Redundancy Check (CRC) \cite{S.Lin} to each transmitted data packet of the source MT. If the decoded information date packet is failed with the CRC, the RN will request the MTs to transmit the punctured coded bits at the MTs or repetition coded bits according to the rate compatible criterion \cite{lin_IT2000}. This ARQ scheme ensures that the codewords of the joint distributed encoders for both MTs can be successfully decoded at the RN.
The bit error probability for a memoryless information source sequence of the distributed network coded modulation system will follow the following theorem.
\begin{theorem}
%\textbf{Theorem:} %{[Upper bound on the symbol error probability
%for a memoryless uniform source]}
Under the MLSD and the assumption that the source block is infinitely long, the bit error probability for a distributed convolutional encoded CPFSK system
with a discrete memoryless uniform digital source sequence, 
can be
upper bounded by Eq. (\ref{upboundsym2}),%\vspace{-2ex}
\begin{figure*}[!t]
\normalsize
\begin{eqnarray}\label{upboundsym2} %\vspace{-2ex}%\tiny
P_b^d&{<}&Q\left(\sqrt{d_{min}^2\frac{E_br}{N_0}}\right)\exp\left(d_{min}^2\frac{E_br}{2N_0}\right){\cdot}%\nonumber\\
 %&&
 \frac{{\partial}F(\eta,\epsilon,\zeta)}{{\partial}\epsilon}\mid_{\eta=1/2,\epsilon=1,\zeta=e^{(-E_br/{2N_0})}},
\end{eqnarray}
\hrulefill
\end{figure*}
where $d_{min}^2$ is the minimum NSED and $\eta$, $\epsilon$,
$\zeta$ are dummy variables and $r$ is the code rate of the joint trellis
encoder of all the source nodes, $r=p/s$ as described in Section II. %$E_s$ is the information symbol energy.
The average transfer function is given by Eq. (\ref{eq:bd1}), %$T(\Gamma,I,W)$ is %calculated
%by
%\vspace{-2ex}
\begin{figure*}[!t]
\normalsize
\begin{eqnarray}\label{eq:bd1}%\vspace{-2ex}%\small
F(\eta,\epsilon,\zeta)&=&M^{-m}\sum_{\kappa=1}^{M^m}F(j,\kappa,\eta,\epsilon,\zeta)%\nonumber\\
=M^{-m}\frac{1}{p}\sum_{j=0}^{p-1}\sum_{\kappa=1}^{M^m}\sum_{\iota}\sum_{\tau}\sum_d{W_{j,s_\kappa,\iota,\tau,d}}\eta^{\iota}\epsilon^{\tau}\zeta^{d^2} \nonumber\\
\end{eqnarray}
\hrulefill
\end{figure*}
where $W_{j,s_\kappa,\iota,\tau,d}$ is the number of error events
that start at time $j$ from state $s_\kappa$, and have NSED $d^2$,
length $\iota$ and total number of symbol errors caused by the
error event given by $\tau$. The $Q$ function is defined as
$Q(x)=(\sqrt{2\pi})^{-1}\int_{x}^{\infty}e^{-z^2/2}dz$.
\hspace{25ex} $\square$
\end{theorem}
%The proof of the above Theorem is given in the Appendix. 

The transfer function $F(j,\kappa,\eta,\epsilon,\zeta)$ can be obtained by using
a product state diagram \cite{Biglieri,productstate}. A product
state at time $j$ is defined as
$({\mathbf{\boldsymbol{\sigma}}}_j,{\mathbf{\boldsymbol{\hat\sigma}}}_j)$,
where ${\mathbf{\boldsymbol{\sigma}}}_j$ is a state of the super-trellis encoder
of the distributed system and ${\mathbf{\boldsymbol{\hat\sigma}}}_j$
represents a state of the decoder. The transition
$({\mathbf{\boldsymbol{\sigma}}}_j,{\mathbf{\boldsymbol{\hat\sigma}}}_j){\rightarrow}
({\mathbf{\boldsymbol{\sigma}}}_{j+1},{\mathbf{\boldsymbol{\hat\sigma}}}_{j+1})$
is labeled with %\vspace{-1ex}
\begin{equation}\label{eq:labels}%\vspace{-1ex}
 \sum_{\Delta{\tau}}\sum_{\Delta{d^2}}b(\Delta{\tau},\Delta{d^2})\eta{\epsilon^{\Delta{\tau}}}\zeta^{\Delta{d^2}},
\end{equation}
where $\Delta{\tau}$ and $\Delta{d^2}$ are the number of the
symbol errors and NSED, respectively.
$b(\Delta{\tau},\Delta{d^2})$ denotes the number of paths having
NSED $\Delta{d^2}$ and symbol errors $\Delta{\tau}$ for this
state transition.

For the distributed convolutional coded CPFSK system, the bit error rate only depends on the output of
the distributed convolutional encoder. In other words, it is independent
of the pair state of the RSC within the CPFSK modulator
$(\mathbf{\boldsymbol{\sigma}}_j^{ne},\mathbf{\boldsymbol{\hat\sigma}}_j^{ne})$.
Furthermore, the NSED $d^2$ only depends on the difference of the CPFSK
states
$(\mathbf{\boldsymbol{\sigma}}_j^{ne}-\mathbf{\boldsymbol{\hat\sigma}}_j^{ne})$
\cite{LinZihuai_PhDthesis}. Therefore, the product state can be reduced. 
The reduced product
state can be written as
$(\mathbf{\boldsymbol{\sigma}}_j^{cc},\mathbf{\boldsymbol{\hat\sigma}}_j^{cc},\omega_j)$,
where
$\omega_j=\emph{R}_P\left\{(\mathbf{\boldsymbol{\sigma}}_j^{ne}-\mathbf{\boldsymbol{\hat\sigma}}_j^{ne})\right\}=\emph{R}_P\left\{\sum_{n=0}^{j-L}\gamma_n\right\}$
is the difference phase state. %, see (\ref{eq:d2}). %$\emph{R}_x\{\cdot\}$ is the modulo
%$x$ operator.
%$\gamma_n=U_n-\hat{U}_n\in\left\{0,\pm{1},\ldots,\pm{M-1}\right\}$
%is the difference symbol between the transmitted symbol $U_n$ and
%the decoded symbol $\hat{U}_n$ for the CPM.
The total number of product states for the distributed convolutional coded CPFSK systems is $2^{2m+1}$.

The product states can be divided
into initial states, transfer states and end states. A product
state is an initial state if an error event can start from it. A
product state is an end state if an error event can end in it. The
conditions for initial states and end states are
$\mathbf{\boldsymbol{\sigma}}_j^{cc}=\mathbf{\boldsymbol{\hat\sigma}}_j^{cc}$
and $\omega_j=0$. Other states are referred to as transfer states.

Let ${\mathbf{\boldsymbol{A}}}_{\kappa,j}$ represent the state
transitions from an initial state $s_\kappa$ to transfer states in
one step at time $j$. Let us denote by
${\mathbf{\boldsymbol{B}}}_j$ the transitions from transfer states
to end states, and by ${\mathbf{\boldsymbol{C}}}_j$ the
transitions from transfer states to transfer states in one step at
time $j$. Let ${\mathbf{\boldsymbol{D}}}_{\kappa,j}$ represent the
transitions from an initial state $s_\kappa$ to end states in one
step. The transfer function can be calculated by
\cite{IT-CPM}\vspace{-1ex}
\begin{equation}\vspace{-1ex}\label{eq:generatingfunction}
F(j,\kappa,\eta,\epsilon,\zeta)=\mathbf{\boldsymbol{1}}{\cdot}\left({\mathbf{\boldsymbol{B}}}_j({\mathbf{\boldsymbol{I}}}-{\mathbf{\boldsymbol{C}}}_j)^{-1}{\mathbf{\boldsymbol{A}}}_{\kappa,j}+{\mathbf{\boldsymbol{D}}}_{\kappa,j}\right)
\end{equation}
where $\mathbf{\boldsymbol{1}}$ is an all one vector and
$\mathbf{\boldsymbol{I}}$ represents the identity matrix.

Eq. (\ref{upboundsym2}) can be further expressed as
\begin{equation}\label{eq:asymptoticalser}\vspace{-1ex}%\small%\footnotesize
P_b^d {\leq} \sum_{d^2}W_d{\cdot}Q(\sqrt{d^2\frac{E_br}{N_0}}),
\end{equation}
where \vspace{0ex}
%\begin{equation}\label{eq:Cdmin}%\vspace{-1ex}%\small%\footnotesize
$W_d=
\frac{2^{-m}}{p}\sum_{j=1}^p\sum_{\kappa=1}^{2^m}\sum_{l}\sum_{\tau}{W_{j,s_\kappa,l,\tau,d}}{\cdot}{\tau}{\cdot}2^{-l}$.
%\end{equation}

It can be seen from (\ref{eq:asymptoticalser}) that the minimum
NSED $d_{min}^2$ and $W_{d_{min}}$ (the number
of error events with $d_{min}^2$) dominate %Asymptotically (large SNR values),
the asymptotical symbol error rate of the system. %will be
%dominated by .
%\end{itemize}
%
%For a concatenated PRCC/CPM, the NSED $d^2$ associated with an
%error event $\Upsilon=\boldsymbol{\mu}- \boldsymbol{\mu}'$ can be
%calculated as
%\begin{equation}\label{eq:d2}\vspace{-0ex}\footnotesize
%R{\cdot}\log_2M{\cdot}
%(l-\frac{1}{T}\sum_{i=0}^{l-1}\int_{iT}^{(i+1)T}\cos[2{\pi}h\omega_i+4{\pi}h\sum_{j=i-L+1}^{i}\gamma_iq(t-jT)]dt)
%\end{equation}
%\begin{equation*}\label{eq:d2}\vspace{-0ex}
%d^2=\frac{1}{2E_b}\int_{-\infty}^\infty[s(t,\textbf{U})-s(t,\hat{\textbf{U}})]^2dt
%\end{equation*}.
%\begin{itemize}
%\item
%\begin{equation}\label{eq:d2}%\vspace{-0ex}
%d^2=r{\cdot}\log_2M{\cdot}\left(\iota-\frac{1}{T}\sum_{i=0}^{\iota-1}\int_{iT}^{(i+1)T}\cos
%\phi(t,{\boldsymbol{\gamma}})dt\right).
%\end{equation}
%Here $T$ is the symbol interval duration and
%$\phi(t,{\boldsymbol{\gamma}})=[2{\pi}h\omega_i+4{\pi}h\sum_{j=i-L+1}^{i}\gamma_iq(t-jT)]$,
%where
%$\omega_j=\emph{R}_P\left\{({\boldsymbol{\sigma}}_j^{cpm}-{\boldsymbol{\hat\sigma}}_j^{cpm})\right\}=\emph{R}_P\left\{\sum_{n=0}^{j-L}\gamma_n\right\}$
%is the difference phase state
%%\end{equation}
%%$\omega_i=\emph{R}_P\left\{\sum_{n=0}^{i-L}\gamma_n\right\}$ is the
%%difference phase state
%and $r$ is the code rate of the trellis
%encoder of the PRCC, $r=p/s$. %For this case, the source
%%encoding rate $R=\log_2M$ bits/sample.
%$\emph{R}_x\{\cdot\}$ is the modulo $x$ operator and $q(\cdot)$ is
%the phase response \cite{anderson}.

%Largest $d_{min}^2$ is used as the design criterion for the
%selection of the best puncture matrices of the investigated
%distributed network coded systems.

Based on (\ref{eq:BERoverall}), (\ref{eq:BERoverall_2}) and (\ref{upboundsym2}), we can get the overall bit error probability for the investigated distributed network coded modulation system via eq. (\ref{eq:BERoverall_3}).
\begin{figure*}[!t]
\normalsize
\begin{eqnarray}\label{eq:BERoverall_3}
P_b&=&1-(1-P_b^u)(1-P_b^d)=1-(1-(1-Q(\sqrt{rE_b/N_0}))^{N_s})(1-P_b^d) \nonumber\\
&\leq& 1-(1-(1-Q(\sqrt{rE_b/N_0}))^{N_s})(1-Q\left(\sqrt{d_{min}^2\frac{E_br}{N_0}}\right)\exp\left(d_{min}^2\frac{E_br}{2N_0}\right){\cdot}%\nonumber\\
 %&&
 \frac{{\partial}F(\eta,\epsilon,\zeta)}{{\partial}\epsilon}\mid_{\eta=1/2,\epsilon=1,\zeta=e^{(-E_br/{2N_0})}})\nonumber\\
&\leq& 1-\left(1-(1-Q(\sqrt{rE_b/N_0}))^{N_s}\right)\left(1-\sum_{d^2}W_d{\cdot}Q(\sqrt{d^2\frac{E_br}{N_0}})\right)
\end{eqnarray}
\hrulefill
\end{figure*}

\section{Code Search}
To achieve the network coded modulation scheme with nested codes, the code design is assumed to satisfy the following criterion:
\begin{enumerate}
  \item the generators assigned to different nodes should be mutually linearly independent.
  \item the rate of ``stacked'' generator matrix should be less than $1$.
  \item the selected code should not be a catastrophic convolutional code.
\end{enumerate}

\begin{table}[!t]
    \caption{Table of good codes}\label{tab:goodcodes}
    \centering
    \begin{threeparttable}
      \begin{tabular}{*4{c}} %\toprule
      \hline
        Rate & Memory number & Generator matrices &  $d_{free}$ \\ %\midrule
        \hline
        2/3 & 2 & $\left( \begin{array}{ccc}
        6\hspace{2mm} 5 \hspace{2mm}1 \\
        7\hspace{2mm} 2 \hspace{2mm}5
        \end{array} \right)_{8}$ & 5 \\

        2/4 & 2 & $\left( \begin{array}{cccc}
        3\hspace{2mm} 7 \hspace{2mm}1 \hspace{2mm}6 \\
        4\hspace{2mm} 7 \hspace{2mm}6 \hspace{2mm}3
        \end{array} \right)_{8}$  & 8 \\

        3/4 & 2 & $\left( \begin{array}{cccc}
        5\hspace{2mm} 4 \hspace{2mm}3 \hspace{2mm}2 \\
        4\hspace{2mm} 6 \hspace{2mm}5 \hspace{2mm}5 \\
        6\hspace{2mm} 1 \hspace{2mm}4 \hspace{2mm}3
        \end{array} \right)_{8}$  & 6 \\

        4/6 & 2 & $\left( \begin{array}{cccccc}
        5\hspace{2mm} 6 \hspace{2mm}5 \hspace{2mm}6 \hspace{2mm}7 \hspace{2mm}4\\
        7\hspace{2mm} 0 \hspace{2mm}7 \hspace{2mm}3 \hspace{2mm}6 \hspace{2mm}2\\
        4\hspace{2mm} 5 \hspace{2mm}2 \hspace{2mm}6 \hspace{2mm}5 \hspace{2mm}0\\
        6\hspace{2mm} 1 \hspace{2mm}5 \hspace{2mm}7 \hspace{2mm}2 \hspace{2mm}5
        \end{array} \right)_{8}$ & 8 \\ %\bottomrule
        \hline
      \end{tabular}
%      \begin{tablenotes}
%        \footnotesize
%        \item[$\star$]  $d_{free}$ denotes the free distance of a binary convolutional code.
%      \end{tablenotes}
    \end{threeparttable}
\end{table}
We construct several good codes based on the modified FAST algorithm in \cite{12Chang}, which are presented in Table \ref{tab:goodcodes}. Then, we can choose different rows of one generator matrix as different linearly independent generators.
In this work, we choose a rate $2/3$ code from Table I. Table II illustrates the code performance of the selected $2/3$ code. Then, one of the possible strategies to assign the generators to different nodes can be

\begin{eqnarray}
%\begin{split}
 &G_{1} =  \left[ 6\hspace{2mm} 5 \hspace{2mm}1\right]_{8}, \\% \hspace{2mm}6 \hspace{2mm}7 \hspace{2mm}4\right]_{8},\\
 &G_{2} =  \left[7\hspace{2mm} 2 \hspace{2mm}5\right]_{8}. %,\\% \hspace{2mm}3 \hspace{2mm}6 \hspace{2mm}2\right]_{8},\\
 %&G_{C} =  \left[4\hspace{2mm} 5 \hspace{2mm}2 \hspace{2mm}6 \hspace{2mm}5 \hspace{2mm}0\right]_{8},\\
 %&G_{D} =  \left[6\hspace{2mm} 1 \hspace{2mm}5 \hspace{2mm}7 \hspace{2mm}2 \hspace{2mm}5\right]_{8}.
%\end{split}
%\end{equation*}
\end{eqnarray}
\begin{table*}
    \caption{The Analysis of Code Performance}\label{tab:tablenotes}
    \centering
    \begin{threeparttable}
      \begin{tabular}{|c|c|c|c|c|c|c|c|c|c|c|}
      \hline
        \multicolumn{1}{|c|}{\multirow{2}{*}{Rate}} & \multicolumn{1}{|c|}{\multirow{2}{*}{Generator matrices }} &
        \multicolumn{1}{|c|}{\multirow{2}{*}{$d_{free}$}}  & \multicolumn{7}{|c|}{$\frac{c_\varphi|_{\varphi = (d_{free}+i)}}{a_{\varphi}|_{\varphi =(d_{free}+i)}}$}\\\cline{4-10}
        \multicolumn{1}{|c|}{} & \multicolumn{1}{|c|}{} & \multicolumn{1}{|c|}{} &
        $i=0$ & $i=1$ & $i=2$ & $i=3$ & $i=4$ & $i=5$ & $i=6$\\ \hline

        \multicolumn{1}{|c|}{\multirow{2}{*}{1/3}} &
        $\left(\begin{array}{cccccc}
        6 \hspace{2mm} 5 \hspace{2mm} 1 \end{array}\right)_{8}$ & 5 & $\frac{1}{1}$ & $\frac{0}{0}$ & $\frac{0}{0}$ & $\frac{4}{2}$ & $\frac{3}{1}$ & $\frac{4}{1}$ & $\frac{14}{4}$ \\ \cline{2-10}

        \multicolumn{1}{|c|}{} &
        $\left(\begin{array}{cccccc}
        7 \hspace{2mm} 2 \hspace{2mm} 5 \end{array}\right)_{8}$ & 6 & $\frac{1}{1}$ & $\frac{0}{0}$ & $\frac{4}{2}$ & $\frac{0}{0}$ & $\frac{12}{4}$ & $\frac{0}{0}$ & $\frac{32}{8}$ \\ \hline

        \multicolumn{1}{|c|}{\multirow{1}{*}{2/3}} &
        $\left(\begin{array}{cccccc}
        6 \hspace{2mm} 5 \hspace{2mm} 1   \\
        7 \hspace{2mm} 2 \hspace{2mm} 5
        \end{array}\right)_{8}$ & 5 & $\frac{5}{3}$ & $\frac{45}{11}$ & $\frac{218}{39}$ & $\frac{949}{135}$ & $\frac{4518}{519}$ & $\frac{19355}{1902}$ & $\frac{81065}{6875}$  \\\hline
      \end{tabular}
    \end{threeparttable}
  \end{table*}

\section{Simulation Results for Distributed Network Coded Systems}\label{sec:jscnumericalresults}
\begin{figure}[htbp]
\centering{\includegraphics[width = 87 mm]{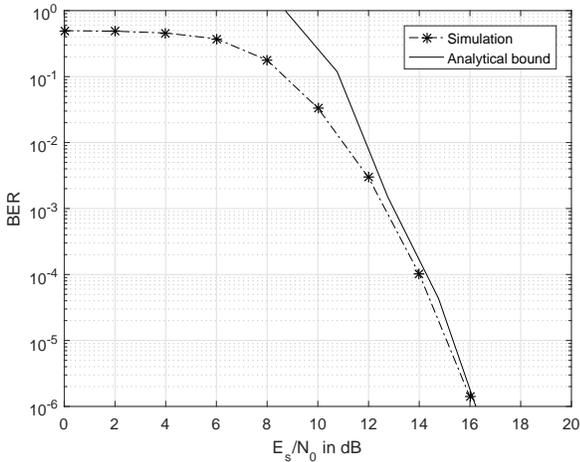}}
\caption{Analytical upper bounds and one simulation result for the investigated systems.}
\label{fig:BERbound_PRCCCPM}
\end{figure}

%\subsection{Simulation results for Distributed Network Coded System}
Computer simulations have been performed for the distributed
physical layer network coded and digital phase modulated systems
over Rayleigh fading and AWGN channels. We consider two mobile stations communicate with BS via a single relay node. The two mobile stations are encoded with 
the generator polynomial matrices
$G_1(D)=[1+D;1+D^2;D^2]$, $G_2(D)=[1+D+D^2;D;1+D^2]$, respectively. 
The mobile stations transmit the information packets via orthogonal channels. When the relay received the codewords from the mobile stations, it began to decode. The decoded codewords are then network encoded, and fed into the NE and transmitted to the BS after the CPFSK modulation. In the simulation, we consider the channels between the mobile stations and relay node and between the relay node and the BS are AWGN channels. 
For each channel SNR
value, $1000$ source blocks each with $1000$ bits were used in the
simulation. Fig. \ref{fig:BERbound_PRCCCPM}  shows the upper bound and the simulation
results of BER performance for the investigated coded systems. It can be
seen that the simulation results agree with the upper bound especially when the channel SNR increases. %It also shows that the
%error probability performance for parent codes is better than the punctured codes.  For example, for $G=[1;D+D^2]$, at BER of $10^{-4}$, %When the channel SNR is larger than $\sim 4.8$
%%dB,
%a performance gain of roughly $\sim 4$ dB can be obtained
%over the rate $3/4$ punctured coded system.

%\begin{figure}[htb]
%\begin{center}
%\mbox{\epsfig{file=sim_rate23_codedCPM_results.eps,width=160mm}}
%D:/work/Licproject/vq/tcq/ringTCQ/codedcpmringcc/sim_rate23_codedCPM_AWGNv5_4.eps,width=160mm
%{\epsfig{file=C:/MATLAB6p5/work/Licproject/vq/tcq/tcqcpm/itercpm/bitinterleaver/Hconnection/finalresults/R2H110204M81RECh13vs1525BL403ExnewvsTCMpskvsSCTCQMOPnewTused.eps,width=90mm}}\quad
%\caption{Analytical upper bounds and one simulation result for the investigated systems.
%The corresponding source encoding rates are  $R=2,2\frac{1}{3}, 2\frac{2}{3},3, 3\frac{1}{3}$ bits per source sample.
%} \label{fig:BERbound_PRCCCPM}%}
%\vspace{-3ex}
%\end{center}
%\end{figure}

\section{Summary}\label{sec:jscconclusion}

In this work, we proposed a distributed network coded modulation scheme for multiple access channel in a cellular system. We considered a system in which a number of source MTs transmit information data packets to a BS via a relay node. %In the cooperation phase, the source MT and the cooperative MT send the same data packet to the relay node using their own encoder via orthogonal channels. At the relay node, the received waveforms from both the MTs were symbol wise alternatively concatenated, then demodulated and decoded. There was an ARQ mechanism to ensure reliable transmissions from the MTs to the relay.

The decoded codewords from the MTs at the relay were then fed into a network coded modulator to perform network encoding and digital phase coded modulation. %In the first scheme which we investigated, the correctly decoded codewords from both the MTS were directly fed into the network coded modulator. While in the second scheme, the decoded codeword was first interleaved then fed into the network coded modulator.
Analytical bound on BER performance for the proposed distributed network coded modulation systems
is derived. The bounds are shown asymptotically tight. These bounds can be served as the guideline for the design of the investigated distributed network coded modulation systems.

\end{document}